\documentclass[conference]{IEEEtran}
\IEEEoverridecommandlockouts

\usepackage{cite}
\usepackage{amsmath,amssymb,amsfonts}
\usepackage{lipsum}
\usepackage{algorithmic}
\usepackage[colorinlistoftodos]{todonotes}
\usepackage{graphicx}
\usepackage{textcomp}
\usepackage{acronym}
\usepackage{xcolor}
\def\BibTeX{{\rm B\kern-.05em{\sc i\kern-.025em b}\kern-.08em
    T\kern-.1667em\lower.7ex\hbox{E}\kern-.125emX}}
\usepackage[normalem]{ulem}

\usepackage{tikz}
\usepackage{graphicx}
\usetikzlibrary{positioning, arrows.meta, calc}
\usepackage{circuitikz}
\usepackage{siunitx}
\usepackage{booktabs}
\usepackage{multirow}
\usepackage[
    hidelinks=false,
]{hyperref} 
\usepackage[acronym]{glossaries}
\makeglossaries
\glsdisablehyper

\definecolor{light_r}{HTML}{EC9A9A}
\definecolor{light_b}{HTML}{8BC1F7}
\definecolor{light_g}{HTML}{80ce87}
\definecolor{light_o}{HTML}{FFD580}
\definecolor{my_red}{HTML}{ff2a2a}
\definecolor{my_blue}{HTML}{0066cc}
\definecolor{my_green}{HTML}{008121}
\definecolor{my_orange}{HTML}{FFA500}

\DeclareSIUnit{\pu}{pu}

\newcommand{\arrowSize}{1.6mm}
\newcommand{\measArrowLength}{0.5cm}
\newcommand{\shortBipoleLength}{0.6cm}

\ctikzset{bipoles/length=1cm}

\newcommand{\busvertical}[1]{%
    coordinate(tmp)
    ++(0,0.25cm) edge[line width=1.5pt] ++(0,-0.5cm) node[above] {#1};
    (tmp)
}

\newcommand\addshortterm[2]{
    \newglossaryentry{#1}{name ={#1}, description={#2}, first ={#2 (#1)}, firstplural={#2s (#1s)}}
}
\newcommand\addterm[2]{
    \newglossaryentry {#1}{type=\acronymtype, name={#1}, description={#2}, first={#2}, text={#2}}
}

\addterm{AEMO}{Australian Energy Market Operator}
\addterm{AVR}{automatic voltage regulator}
\addshortterm{CC-GFMI}{current-controlled grid-forming inverter}
\addshortterm{EMT}{electromagnetic transient}
\addshortterm{GFLI}{grid-following inverter}
\addshortterm{GFMI}{grid-forming inverter}
\addshortterm{IBR}{inverter-based resource}
\addshortterm{PCC}{point of common coupling}
\addshortterm{PLL}{phase-locked loop}
\addterm{PWM}{pulse width modulation}
\addterm{PSS}{power system stabilizer}
\addshortterm{RMS}{root-mean-square}
\addterm{SCR}{short circuit ratio}
\addshortterm{SG}{synchronous generator}
\addshortterm{SMIB}{single machine infinite bus}
\addshortterm{SV}{singular value}
\addterm{SVD}{singular value decomposition}
\addshortterm{VC-GFMI}{voltage-controlled grid-forming inverter}
\addterm{VSC}{voltage-sourced converter}
\addterm{VSM}{virtual synchronous machine}

\newcommand\dqVector[1]{
    \boldsymbol{#1}
}

\newcommand\dqVectorSub[2]{
    \boldsymbol{#1}_{#2}
}

\newcommand\dqMatrix[1]{
    \boldsymbol{#1}_{dq}
}
\newcommand\dqMatrixSup[2]{
    \boldsymbol{#1}_{dq}^\mathrm{#2}
}

\newcommand\refValue[1]{
    {#1}^\mathrm{ref}
}

\newcommand\lineParameter[1]{
    {#1}_{l}
}
\newcommand\trafoParameter[1]{
    {#1}_{t}
}
\newcommand\theveninParameter[1]{
    {#1}_\mathrm{th}
}
\newcommand\loadParameter[1]{
    L_{#1}
}
\newcommand\filterParameter[1]{
    {#1}_f
}

\begin{document}

\title{Comparative Assessment of Frequency Scans\\using EMT and RMS Models}

\author{\IEEEauthorblockN{Clea Bürgel, Gustavo Valverde, and Gabriela Hug}
\IEEEauthorblockA{\textit{Power Systems Laboratory} \\
\textit{ETH Zürich}\\
Zürich, Switzerland \\
\{cbuergel, gustavov, hug\}@ethz.ch}
}

\maketitle

\begin{abstract}
Frequency scans of inverter and grid impedances are crucial for assessing small-signal stability and system strength in inverter-dominated power systems. This paper compares frequency-scan results for several generating units, including a synchronous generator, grid-following, and grid-forming inverters, using fully \textit{dq} electromagnetic transient and root-mean-square (RMS) models in Simulink. The scans are performed for the same device rating and operating point. Frequency scans and time-domain simulations show that the RMS models are valid only at low frequencies. We find that the current-controlled grid-forming inverter is more difficult to represent in RMS than the voltage-controlled grid-forming inverter. However, including the inner voltage-control loops improves the accuracy of the RMS representation. The paper also investigates the influence of generator and grid impedance on the total system impedance seen from the point of common coupling and relates the oscillatory modes and zeros of the system to peaks and dips in the first and second singular values of the system impedance, respectively.

\end{abstract}


\section{Introduction}
\label{sec:intro}

The integration of renewable generation and the decommissioning of coal-, nuclear-, and gas-fired power plants are fundamentally changing the way power systems are operated, controlled, and analyzed. The fast and slow interactions of \glspl{IBR} with the network and other system components span both electromagnetic and electromechanical phenomena, demanding more sophisticated simulation frameworks~\cite{RevisitingPowerSystems2024Lara}.

As \glspl{IBR} increasingly drive system dynamics, \gls{RMS} simulations, intended to assess electromechanical phenomena, need to be replaced by high-fidelity \gls{EMT} simulations capable of capturing a broader frequency range. However, the high computational cost of \gls{EMT} simulations has hindered their widespread adoption, especially for large-scale power systems, and many system operators continue to rely on \gls{RMS}-based models and simulations. This inevitably needs to change as \gls{IBR} penetration grows, rendering \gls{EMT} the standard planning tool for assessing system stability.

Frequency scans of inverter and grid impedance at the \gls{PCC} have been adopted to study potential dynamic interactions, identify resonance frequencies, as well as support the assessment of impedance-based small-signal stability with Bode plots and the Nyquist criterion~\cite{EMTbaseddynamic2026Jiang, ImpedanceBasedStabilityCriterion2011Sun}. Additionally, they are used to assess frequency-dependent system strength beyond the fundamental frequency~\cite{ImpedanceMarginRatio2024Zhu,FrequencyAveragedGridImpedanceLamrani}. The \gls{AEMO} defines system strength as the ability of the power system to maintain and control the voltage waveform at any given location in the system \cite{systemstrengthexplained}. Consequently, in a strong system, the voltage sensitivity to changes in current should be low, a property described by the frequency-dependent system impedance.

While previous works have compared frequency scans of \gls{IBR}~\cite{HowManyGridForming2025Xin} and \gls{SG} \cite{ImpedanceScan2024Chakraborty} impedances, there is limited work addressing the differences between results obtained from EMT and RMS models. This paper presents a systematic comparison of the frequency scans of a \gls{SG}, a \gls{GFLI}, and two variants of \glspl{GFMI} with the same power rating, step-up transformer, and operating point using \gls{EMT} and \gls{RMS} models. We investigate the limitations of \gls{RMS} models through frequency scans and identify which type of \gls{GFMI} can be more accurately represented by an \gls{RMS} model in the low-frequency range. Finally, we analyze the system impedance at the \gls{PCC} and clarify how the individual generator and grid impedances shape the resulting frequency-dependent system impedance. These results offer a basis for a clearer interpretation of PCC frequency scans.

The remainder of this paper is organized as follows: Section~\ref{sec:methodology} introduces the methodology and models used to compute the frequency scans of the different devices and the grid. Section~\ref{sec:results} describes the test system and the main simulation results. Finally, conclusions and future work are presented in Section~\ref{sec:conclusions}.

\section{Methodology}
\label{sec:methodology}

\subsection{Frequency Scan}
\label{sec:Impedance_Scan}
The frequency-dependent system impedance in the \textit{dq} reference frame is obtained by a model-driven method \cite{ImpedanceMarginRatio2024Zhu}. Assuming a white-box model of the power system under study, we linearize the non-linear system equations around a given operating point to obtain the state-space matrices $\boldsymbol{A},\, \boldsymbol{B},\, \boldsymbol{C},\, \boldsymbol{D}$. The state-space matrices are then transformed to the transfer function $\boldsymbol{G}(s)$. By selecting current perturbations $\Delta\dqVectorSub{i}{g}(s)$ as inputs and voltages $\Delta\dqVectorSub{u}{j}(s)$ as outputs, the transfer function $\boldsymbol{G}(s)$ corresponds to the system impedance $\dqMatrix{Z}(s)$ seen from bus $j$:
\begin{align}
    \boldsymbol{G}(s) = \boldsymbol{C}(s\boldsymbol{I}-\boldsymbol{A})^{-1}\boldsymbol{B}+\boldsymbol{D} = \dqMatrix{Z}(s) \, .
\end{align}
Letting $s=\mathrm{j}\omega$ yields the frequency scan of $\dqMatrix{Z}(s)$.
Since this impedance is a 2-by-2 matrix at each frequency, the voltage response to current perturbations depends on the perturbation direction in the \textit{dq} plane and therefore cannot be characterized by a single scalar. To address this, we apply the \gls{SVD}~\cite{QuantifyingGridFormingBehavior2025Zhuang, HowManyGridForming2025Xin}. The first and second \glspl{SV} of the impedance matrix, $\sigma_1(\dqMatrix{Z}(s))$ and $\sigma_2(\dqMatrix{Z}(s))$, provide the upper and lower bounds, respectively, on the gain from current perturbations to voltage perturbations:
\begin{equation}
    \sigma_2(\dqMatrix{Z}(s)) \leq \frac{\lVert\Delta\dqVectorSub{u}{j}(s)\rVert_2}{\lVert\Delta\dqVectorSub{i}{g}(s)\rVert_2} \leq  \sigma_1(\dqMatrix{Z}(s)) \, .
\end{equation}
We use frequency scans of these \glspl{SV} to characterize the system's dynamic behavior. Hereafter, we refer to them simply as frequency scans.
Since the first \gls{SV} captures the worst-case amplification, it can also be used as a metric for system strength \cite{QuantifyingGridFormingBehavior2025Zhuang}. The lower the first \gls{SV}, the stronger the system.
In addition to determining the system impedance $\dqMatrix{Z}(s)$ at a given bus, this technique enables the identification of the generator impedance $\dqMatrixSup{Z}{gen}(s)$ and the grid impedance $\dqMatrixSup{Z}{grid}(s)$ at the same bus. The dependency between system, generator, and grid impedances can be derived as~\cite{EMTbaseddynamic2026Jiang}:
\begin{align}
    \dqMatrix{Z}(s) = \left(\boldsymbol{I}+\dqMatrixSup{Z}{gen}(s) \dqMatrixSup{Z}{grid}(s)^{-1}\right)^{-1} \dqMatrixSup{Z}{gen}(s) \, .
\end{align}
Hence, for a large $\dqMatrixSup{Z}{gen}(s)$, we have $\dqMatrix{Z}(s) \approx \dqMatrixSup{Z}{grid}(s)$, and for a large $\dqMatrixSup{Z}{grid}(s)$, we have $\dqMatrix{Z}(s) \approx \dqMatrixSup{Z}{gen}(s)$. This is equivalent to two impedances connected in parallel, where the overall behavior is governed by the smaller impedance.

\subsection{Generator Models}
The various \gls{EMT} models, along with the simplifications for the \gls{RMS} models, are described below.

\subsubsection{Synchronous Generator}
We adopt the eighth-order model described in \cite[(5.16-1)-(5.16-19)]{Analysiselectricmachinery2002Krause}, which represents the fast dynamics of the stator windings, the field winding, and three damper windings, as well as the slower dynamics governed by the rotor-motion equations. The machine model is complemented by a second-order turbine model, a first-order \gls{AVR} model, and a tuned \gls{PSS} to damp electromechanical oscillations. 

\subsubsection{Inverter-Based Resources}
For the \glspl{IBR}, we consider a \gls{VSC} with an $LC$~filter connected to a step-up transformer. We assume an ideal DC source and neglect the switching dynamics. The \gls{GFLI} and \gls{CC-GFMI} models are shown in Fig.~\ref{fig:GFL_Model} and Fig.~\ref{fig:GFM_Model}, respectively.

The \gls{GFLI} controls its current $\dqVector{i}$ and synchronizes to the grid through a \gls{PLL}~\cite{ExaminingFeasibilityModeling2023Mohiuddin}.
The outer control loops~\cite{Gridfollowingconverters2023Lamrani} provide the current reference $\refValue{\dqVector{i}}$ for the inner current-control loops required to track the active power $\refValue{P}$ and voltage setpoints $\refValue{U}$. The inner current-control loops~\cite{AutomaticTuningCascaded2015DArco} control the current $\dqVector{i}$ and generate the required internal voltage command $\refValue{\dqVector{v}}$. 

The \gls{CC-GFMI} imposes its own voltage waveform and indirectly controls the terminal voltage $\dqVector{u}$. The active power control loop, in the considered case a \gls{VSM} \cite{VirtualSynchronousMachine2015DArco}, determines the terminal voltage angle $\theta$, and the outer voltage control loop \cite{Gridfollowingconverters2023Lamrani} provides the terminal voltage reference $\refValue{u}_d$ to track the voltage setpoint. The inner voltage-control loops \cite{AutomaticTuningCascaded2015DArco} output the reference current required to track this terminal voltage reference. The inner current-control loops are the same as for the \gls{GFLI}.

The \gls{VC-GFMI} is considered in Section~\ref{sec:EMTvsRMS} only, since the \gls{CC-GFMI} is more commonly adopted \cite{InvestigationBenedetti2024}. Modeling details for the \gls{VC-GFMI} are found in \cite[single-loop controls]{ComparativeModelingAnalysis2024Favuzza}.

\begin{figure}[t]
\centering
    \begin{circuitikz}[
        auto, node distance=1cm, >=latex, american, font=\footnotesize,
        myArrow/.style={-{Latex[length=\arrowSize, width=\arrowSize]}},
        myArrow_inv/.style={{Latex[length=\arrowSize, width=\arrowSize]}-},
        myCurrentArrow/.style={-{Latex[length=1.25mm, width=1.25mm]}},
        myMeasArrow/.style={-{Latex[length=\arrowSize, width=\arrowSize]}, dashed},
        nodeStyle/.style={draw, rectangle, rounded corners, align=center, minimum width=1.8cm, minimum height=1.2cm},
        nodeStyleSmall/.style={draw, rectangle, rounded corners, align=center, minimum width=1.2cm, minimum height=1.2cm}
    ]

    \node [nodeStyleSmall] (VSC) {\includegraphics[width=0.9cm]{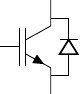}};

    \draw (VSC.east) -- ++(0.5cm,0) coordinate(v) \busvertical{$\dqVector{v}$};
    \draw (v) -- ++(0.5cm,0) to[R, l=$\filterParameter{R}$] ++(1cm,0) -- ++(0.25cm,0) coordinate(RL);
    \draw (RL) -- ++(0.25cm,0) to[L, l=$\filterParameter{L}$] ++(1cm,0) -- ++(1.8cm,0) coordinate(LC);
    \draw (LC) to[C, l=$\filterParameter{C}$] ++(0,-1cm) node[tlground](Ground){};
    \draw (LC) -- ++(1.8cm,0) coordinate(u)  \busvertical{$\dqVector{u}$};

    \draw[myMeasArrow] ($(LC)-(1.2cm,-1mm)$) -- ($(LC)+(-1.2cm,\measArrowLength)$) node[above]{$\dqVector{i}$};
    \draw ($(LC)-(1.2cm,0)$) ellipse (0.5mm and 1mm);
    \draw[myMeasArrow] ($(LC)-(0.6cm,0)$) -- ($(LC)+(-0.6cm,\measArrowLength)$) node[above]{$\dqVector{u}$};
    \fill ($(LC)-(0.6,0)$) circle (0.6mm);
    \draw[myMeasArrow] ($(LC)+(0.9cm,1mm)$) -- ($(LC)+(0.9cm,\measArrowLength)$) node[above, yshift=-0.75mm]{$\dqVectorSub{i}{g}$};
    \draw ($(LC)+(0.9cm,0)$) ellipse (0.5mm and 1mm);
    
    \node [nodeStyle, below=1.5cm of LC, xshift=-0.2cm] (PLL) {Phase-locked\\loop};
    \node [nodeStyle, left=1.8cm of PLL] (CC) {Inner current\\control};
    \node [nodeStyle, below=0.75cm of CC] (OL) {Outer control\\loops};
    
    \draw[myArrow] ($(PLL.west)-(0,0.25cm)$) -- ($(CC.east)-(0,0.25cm)$);
    \node[above right, xshift=\measArrowLength] at ($(CC.east)-(0,0.25cm)$) {$\omega$};
    \draw[myArrow] ($(PLL.west)+(0,0.25cm)$) -- ($(PLL.west)+(-0.5cm,0.25cm)$) node[left]{$\theta$};
    \draw[myMeasArrow] ($(PLL.east)+(\measArrowLength,0)$) node[right]{$u_{q}$} -- (PLL.east);
    \draw[myArrow] ($(PLL.south)-(0, \measArrowLength)$) node[below]{$\refValue{u}_q=0$} -- (PLL.south);

    \draw[myArrow] (OL) -- (CC) node[midway, right]{$\refValue{\dqVector{i}}$};
    \draw[myArrow, rounded corners] (CC) -- (CC -| VSC) node[midway, above]{$\refValue{\dqVector{v}}$} -- (VSC);
    \draw[myMeasArrow] ($(CC.north)+(-0.3cm,\measArrowLength)$) node[above]{$\dqVector{i}$} -- ($(CC.north)+(-0.3cm,0)$);
    \draw[myMeasArrow] ($(CC.north)+(0.3cm,\measArrowLength)$) node[above]{$\dqVector{u}$} -- ($(CC.north)+(0.3cm,0)$);

    \draw[myArrow] ($(OL.west)+(-\measArrowLength,0.25cm)$) node[left]{$\refValue{P}$} -- ($(OL.west)+(0,0.25cm)$);
    \draw[myArrow] ($(OL.west)+(-\measArrowLength,-0.25cm)$) node[left]{$\refValue{U}$} -- ($(OL.west)+(0,-0.25cm)$);
    \draw[myMeasArrow] ($(OL.east)-(-\measArrowLength,0.25cm)$) node[right]{$\dqVector{u}$} -- ($(OL.east)-(0,0.25cm)$);
    \draw[myMeasArrow] ($(OL.east)+(\measArrowLength,0.25cm)$) node[right]{$\dqVectorSub{i}{g}$} -- ($(OL.east)+(0,0.25cm)$);

    
    \end{circuitikz}
    \caption{\gls{EMT} \gls{GFLI} model in the \textit{dq} reference frame of the generator.}
    \label{fig:GFL_Model}
\end{figure}
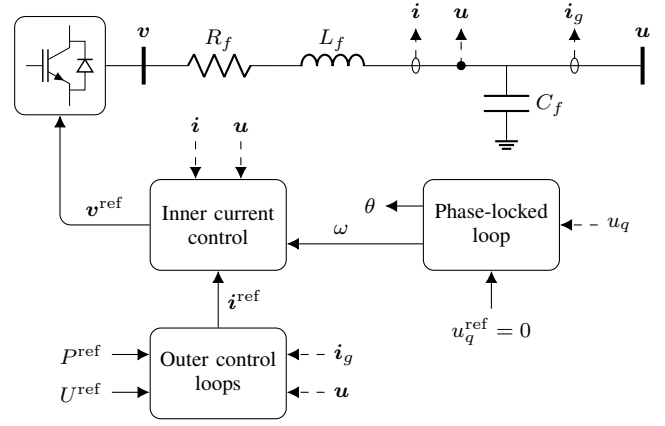
\begin{figure}[t]
\centering
    \begin{circuitikz}[
        auto, node distance=1cm, >=latex, american, font=\footnotesize,
        myArrow/.style={-{Latex[length=\arrowSize, width=\arrowSize]}},
        myArrow_inv/.style={{Latex[length=\arrowSize, width=\arrowSize]}-},
        myCurrentArrow/.style={-{Latex[length=1.25mm, width=1.25mm]}},
        myMeasArrow/.style={-{Latex[length=\arrowSize, width=\arrowSize]}, dashed},
        nodeStyle/.style={draw, rectangle, rounded corners, align=center, minimum width=1.8cm, minimum height=1.2cm},
        nodeStyleSmall/.style={draw, rectangle, rounded corners, align=center, minimum width=1.2cm, minimum height=1.2cm}
    ]

    \node [nodeStyleSmall] (VSC) {\includegraphics[width=0.9cm]{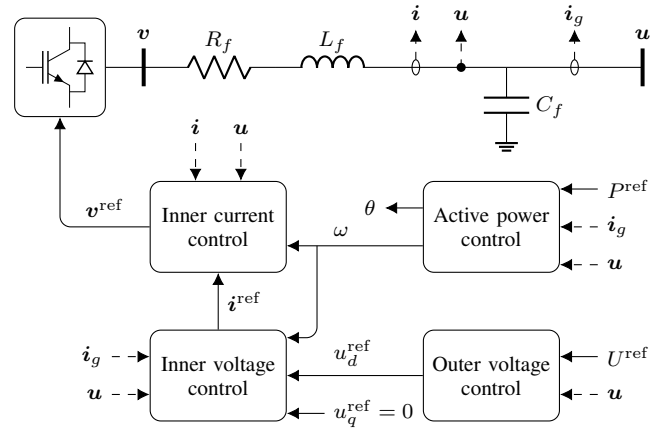}};

    \draw (VSC.east) -- ++(0.5cm,0) coordinate(v) \busvertical{$\dqVector{v}$};
    \draw (v) -- ++(0.5cm,0) to[R, l=$\filterParameter{R}$] ++(1cm,0) -- ++(0.25cm,0) coordinate(RL);
    \draw (RL) -- ++(0.25cm,0) to[L, l=$\filterParameter{L}$] ++(1cm,0) -- ++(1.8cm,0) coordinate(LC);
    \draw (LC) to[C, l=$\filterParameter{C}$] ++(0,-1cm) node[tlground](Ground){};
    \draw (LC) -- ++(1.8cm,0) coordinate(u)  \busvertical{$\dqVector{u}$};

    \draw[myMeasArrow] ($(LC)-(1.2cm,-1mm)$) -- ($(LC)+(-1.2cm,\measArrowLength)$) node[above](a){$\dqVector{i}$};
    \draw ($(LC)-(1.2cm,0)$) ellipse (0.5mm and 1mm);
    \draw[myMeasArrow] ($(LC)-(0.6cm,0)$) -- ($(LC)+(-0.6cm,\measArrowLength)$) node[above]{$\dqVector{u}$};
    \fill ($(LC)-(0.6,0)$) circle (0.6mm);
    \draw[myMeasArrow] ($(LC)+(0.9cm,1mm)$) -- ($(LC)+(0.9cm,\measArrowLength)$) node[above, yshift=-0.75mm](b){$\dqVectorSub{i}{g}$};
    \draw ($(LC)+(0.9cm,0)$) ellipse (0.5mm and 1mm);
    
    \node [nodeStyle, below=1.5cm of LC, xshift=-0.2cm] (VSM) {Active power\\control};
    \node [nodeStyle, below=0.75cm of VSM] (OVC) {Outer voltage\\control};
    \node [nodeStyle, left=1.8cm of VSM] (CC) {Inner current\\control};
    \node [nodeStyle, below=0.75cm of CC] (IVC) {Inner voltage\\control};
    
    \draw[myArrow] ($(VSM.west)-(0,0.25cm)$) -- ($(CC.east)-(0,0.25cm)$) node[midway, above, xshift=-\measArrowLength](w){};
    \node[above right, xshift=\measArrowLength] at ($(CC.east)-(0,0.25cm)$) {$\omega$};
    \draw[myArrow] ($(VSM.west)+(0,0.25cm)$) -- ($(VSM.west)+(-0.5cm,0.25cm)$) node[left]{$\theta$};
    \draw[myArrow] ($(VSM.east)+(\measArrowLength,0.5cm)$) node[right]{$\refValue{P}$} -- ($(VSM.east)+(0,0.5cm)$);
    \draw[myMeasArrow] ($(VSM.east)+(\measArrowLength,0)$) node[right]{$\dqVectorSub{i}{g}$} -- (VSM);
    \draw[myMeasArrow] ($(VSM.east)+(\measArrowLength,-0.5cm)$) node[right]{$\dqVector{u}$} -- ($(VSM.east)+(0,-0.5cm)$);

    \draw[myArrow] (IVC) -- (CC) node[midway, right]{$\refValue{\dqVector{i}}$};
    \draw[myArrow, rounded corners] (CC) -- (CC -| VSC) node[midway, above]{$\refValue{\dqVector{v}}$} -- (VSC);
    \draw[myMeasArrow] ($(CC.north)+(-0.3cm,\measArrowLength)$) node[above]{$\dqVector{i}$} -- ($(CC.north)+(-0.3cm,0)$);
    \draw[myMeasArrow] ($(CC.north)+(0.3cm,\measArrowLength)$) node[above]{$\dqVector{u}$} -- ($(CC.north)+(0.3cm,0)$);
    
    \draw[myArrow] (OVC) -- (IVC);
    \node[above right, xshift=\measArrowLength] at (IVC.east) {$\refValue{u}_d$};
    \draw[myArrow, rounded corners] (w.south) -- ($(w.south |- IVC.east) + (0,5mm)$) -- ($(IVC.east) + (0,5mm)$);
    \draw[myMeasArrow] ($(IVC.west)+(-\measArrowLength,0.25cm)$) node[left]{$\dqVectorSub{i}{g}$} -- ($(IVC.west)+(0,0.25cm)$);
    \draw[myMeasArrow] ($(IVC.west)+(-\measArrowLength,-0.25cm)$) node[left]{$\dqVector{u}$} -- ($(IVC.west)+(0,-0.25cm)$);
    \draw[myArrow] ($(IVC.east)-(-\measArrowLength,0.5cm)$) node[right]{$\refValue{u}_q=0$} -- ($(IVC.east)-(0,0.5cm)$);

    \draw[myArrow] ($(OVC.east)+(\measArrowLength,0.25cm)$) node[right]{$\refValue{U}$} -- ($(OVC.east)+(0,0.25cm)$);
    \draw[myMeasArrow] ($(OVC.east)+(\measArrowLength,-0.25cm)$) node[right]{$\dqVector{u}$} -- ($(OVC.east)+(0,-0.25cm)$);

    
    \end{circuitikz}
    \caption{\gls{EMT} \gls{CC-GFMI} model in the \textit{dq} reference frame of the generator.}
    \label{fig:GFM_Model}
\end{figure}

\subsubsection{Simplifications in RMS}
The following simplifications are made for the \gls{RMS} models~\cite{RevisitingPowerSystems2024Lara}: the fast dynamics of transmission lines, transformers, and filters, as well as the stator dynamics of rotating machines, are neglected, i.e., the corresponding equations are purely algebraic. The fast inner current-control and the inner voltage-control loops of inverters are assumed ideal and are therefore not modeled.

\section{Results}
\label{sec:results}

In this section, we first present the simulation setup and then compare the frequency scans obtained using \gls{EMT} and \gls{RMS} models. Finally, we focus on the \gls{EMT} frequency scans and provide a more detailed interpretation.

\subsection{Simulation Setup}

\begin{figure}[t]
\centering
    \begin{circuitikz}[
        auto, node distance=1cm, >=latex, american, font=\footnotesize,
        myArrow/.style={-{Latex[length=\arrowSize, width=\arrowSize]}},
        myArrow_inv/.style={{Latex[length=\arrowSize, width=\arrowSize]}-},
        myCurrentArrow/.style={-{Latex[length=1.25mm, width=1.25mm]}},
        myMeasArrow/.style={-{Latex[length=\arrowSize, width=\arrowSize]}, dashed},
        myImpedanceArrow/.style={-{Triangle[length=0.5cm, width=0.7cm]}, line width=0.5cm},
        nodeStyle/.style={draw, rectangle, rounded corners, align=center, minimum width=1.8cm, minimum height=1.2cm},
        nodeStyleSmall/.style={draw, rectangle, rounded corners, align=center, minimum width=1cm, minimum height=1.2cm},
        nodeStyleCircle/.style={draw, circle, align=center, minimum width=0.1cm, minimum height=0.1cm}
    ]

    \node[nodeStyleCircle, thick] (Gen) {G};
    \draw (Gen.east) -- ++(0.5cm,0) coordinate(u) \busvertical{$\dqVector{u}$};

    \draw (u) -- ++(0.25cm,0) to[R, l=$\trafoParameter{R}$, bipoles/length=\shortBipoleLength] ++(0.6cm,0) -- ++(0.125cm,0) to[L, l=$\trafoParameter{L}$, bipoles/length=\shortBipoleLength] ++(\shortBipoleLength,0) -- ++(0.25cm,0) coordinate(uj) \busvertical{$\dqVectorSub{u}{j}$};

    \node[below left, color=gray] at ($(uj)-(0,0.25cm)$) {PCC};
    \draw[gray, dashed] (uj) ellipse(2mm and 3.5mm);

    \draw (uj) -- ++(1cm,0) coordinate(Rl) to[R, l=$\lineParameter{R}$, bipoles/length=\shortBipoleLength] ++(\shortBipoleLength,0) -- ++(0.0625cm,0) coordinate(RLmid) -- ++(0.0625cm,0) to[L, l=$\lineParameter{L}$, bipoles/length=\shortBipoleLength] ++(\shortBipoleLength,0) coordinate(Ll) -- ++(1cm,0) coordinate(uk) \busvertical{$\dqVectorSub{u}{k}$};

    \draw ($(Rl) - (0.25cm,0)$) to[C, l=$\frac{\lineParameter{C}}{2}$, bipoles/length=\shortBipoleLength] ++(0,-1.25cm) node[tlground, scale=0.8]{};
    \draw ($(Ll) + (0.25cm,0)$) to[C, l_=$\frac{\lineParameter{C}}{2}$, bipoles/length=\shortBipoleLength] ++(0,-1.25cm) node[tlground, scale=0.8]{};

    \draw[myArrow] ($(uj) + (0,-0.125cm)$) -- ++(0.25cm,0) -- ++(0, -0.75cm) node[below]{$\loadParameter{j}$};
    \draw[myArrow] ($(uk) + (0,-0.125cm)$) -- ++(-0.25cm,0) -- ++(0, -0.75cm) node[below]{$\loadParameter{k}$};

    \draw (uk) -- ++(0.25cm,0) to[R, l=$\theveninParameter{R}$, bipoles/length=\shortBipoleLength] ++(\shortBipoleLength,0) -- ++ (0.125cm,0) to[L, l=$\theveninParameter{L}$, bipoles/length=\shortBipoleLength] ++(\shortBipoleLength,0) -- ++ (0.25cm,0)coordinate(inf);
    \node[nodeStyleCircle, thick, anchor=west] at (inf){$\infty$};

    \draw[myCurrentArrow] ($(u)-(0.3125cm,0)$) -- ($(u)-(0.1875cm,0)$) node[midway, below]{$\dqVectorSub{i}{g}$}; 

    \draw[myArrow, line width=1.5pt, color=my_blue] ($(uj) + (-0.1cm, 0.8cm)$) -- ($(uj) + (-0.1cm, 1.05cm)$) -- ($(uj) + (-0.75cm, 1.05cm)$) node[left, color=my_blue]{$\dqMatrixSup{Z}{gen}$};

    \draw[myArrow, line width=1.5pt, color=my_blue] ($(uj) + (0.1cm, 0.8cm)$) -- ($(uj) + (0.1cm, 1.05cm)$) -- ($(uj) + (0.75cm, 1.05cm)$) node[right, color=my_blue] (Zgrid) {$\dqMatrixSup{Z}{grid}$};

    \node[right, color=my_blue] at ($(Zgrid) + (1.25cm,0cm)$) {$\dqMatrix{Z} = \left(\dqMatrixSup{Z}{gen} \parallel \dqMatrixSup{Z}{grid}\right)$};
    
    \end{circuitikz}
    \caption{One-line diagram of the three-bus system modeled in Simulink.}
    \label{fig:SMIB}
\end{figure}
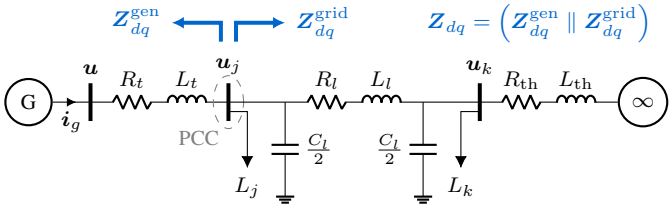

The test system is a radial three-bus system with one generator, two resistive loads, $\loadParameter{j}$ and $\loadParameter{k}$, and an infinite bus behind a Thevenin impedance $\theveninParameter{R}+\mathrm{j}\omega\theveninParameter{L}$ with an X/R ratio of 10, as shown in Fig.~\ref{fig:SMIB}. The generator is connected to the infinite bus through a step-up transformer with resistance $\trafoParameter{R}$ and inductance $\trafoParameter{L}$, and a $\pi$-modeled transmission line with series resistance $\lineParameter{R}$, series inductance $\lineParameter{L}$, and capacitance $\lineParameter{C}$. The generator can be a \gls{SG}, \gls{GFLI}, \gls{CC-GFMI}, or \gls{VC-GFMI}. All system components are custom-developed and modeled in Simulink using \textit{dq} reference frames. At the generator terminals, current and terminal voltage are transformed between the generator's local \textit{dq} reference frame and the grid's synchronous \textit{dq} reference frame.

The operating point is the same for all simulations. The voltage $\dqVectorSub{u}{k}$ is set to $\qty{1}{\pu}$. The generator injects $\qty{0.6}{\pu}$ of active power at unity power factor, and each resistive load consumes \qty{0.15}{\pu} of active power at nominal voltage.

In this work, we investigate the system impedance at the \gls{PCC} $\dqMatrix{Z}(s)$, the generator impedance $\dqMatrixSup{Z}{gen}(s)$, including the generating unit and the step-up transformer, and the grid impedance $\dqMatrixSup{Z}{grid}(s)$, including the loads, the transmission line, and the infinite bus. To obtain these impedances, we specify the corresponding inputs and outputs with \textit{Linear Analysis Points} in the Simulink model, and use the \textit{linearize} command from Simulink Control Design to compute the state-space matrices. The specific input and output settings are given in Table~\ref{tab:io}. To obtain $\dqMatrixSup{Z}{gen}(s)$, the inverse of $\dqMatrixSup{Y}{gen}(s)$ needs to be computed.

\begin{table}[h]
\centering
\caption {Input and output settings for the different scans.}
\label{tab:io}
\begin{tabular}{l|
                l@{\hspace{5pt}}l
                l@{\hspace{5pt}}l}
    \toprule
    Matrix & Inputs & & Outputs &\\
    \midrule
    $\dqMatrix{Z}(s)$ & \textit{Input Perturbations} & $\dqVectorSub{i}{g}$ & \textit{Output Measurements} & $\dqVectorSub{u}{j}$\\[4pt]
    
    $\dqMatrixSup{Y}{gen}(s)$ & \textit{Open-loop Inputs} & $\dqVectorSub{u}{j}$& \textit{Output Measurements} & $\dqVectorSub{i}{g}$\\[4pt]

    $\dqMatrixSup{Z}{grid}(s)$ & \textit{Open-loop Inputs} & $\dqVectorSub{i}{g}$  & \textit{Output Measurements} & $\dqVectorSub{u}{j}$\\
    \bottomrule
\end{tabular}
\end{table}

\subsection{EMT vs. RMS Models}
\label{sec:EMTvsRMS}

\begin{figure}[t]
    \centering
    \includegraphics[width=\linewidth]{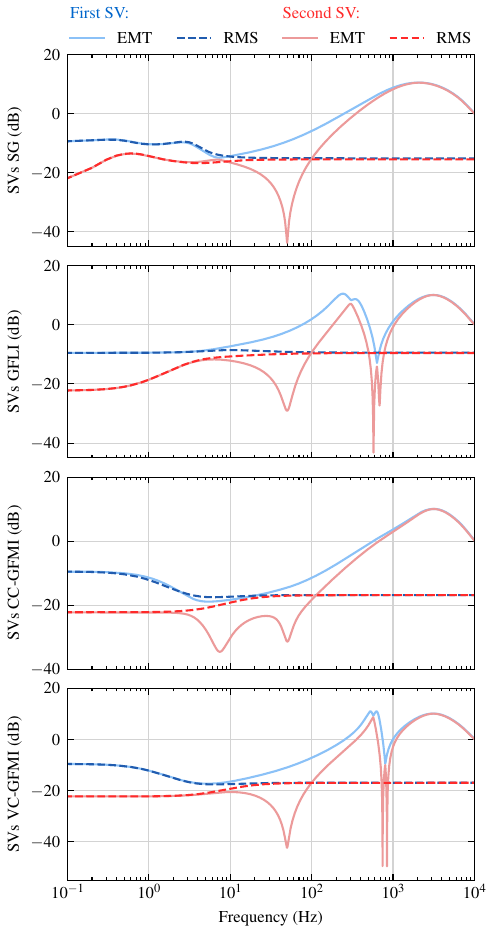}
    \caption{Frequency scans of $\dqMatrix{Z}(s)$ at the \gls{PCC} of the \gls{SG}, \gls{GFLI}, \gls{CC-GFMI}, and \gls{VC-GFMI} using \gls{EMT} and \gls{RMS} models.}
    \label{fig:EMTvsRMS}
\end{figure}

Fig.~\ref{fig:EMTvsRMS} shows the frequency scans of $\dqMatrix{Z}(s)$ at the \gls{PCC} of the \gls{SG}, \gls{GFLI}, \gls{CC-GFMI}, and \gls{VC-GFMI} in the three-bus system, obtained using \gls{EMT} and \gls{RMS} models. The first \gls{SV} of $\dqMatrix{Z}(s)$ determines the maximum voltage deviation in response to a current perturbation, while the second \gls{SV} captures the minimum deviation. Together, they define a frequency-dependent range of possible voltage changes resulting from a current perturbation in any direction of the $dq$ frame. Therefore, when the first and second \glspl{SV} are equal, the voltage deviation is independent of the current perturbation. 

From Fig.~\ref{fig:EMTvsRMS}, we note that all \gls{EMT} scans are similar at very low and very high frequencies due to the grid-impedance dominance. The RMS models can capture the low-frequency dynamics up to approximately \qty{10}{Hz}, but at higher frequencies, the first and second \glspl{SV} converge to the same values, thereby hiding potential high-frequency interactions. This result aligns with the recommendation in \cite{RevisitingPowerSystems2024Lara} that \gls{RMS} simulation should not be used to model dynamics $>\qty{10}{Hz}$.

Analyzing the scans of the two \glspl{GFMI} at low frequencies only, we observe that the \gls{RMS} scan matches the \gls{EMT} scan exactly for the \gls{VC-GFMI}, but small deviations arise for the \gls{CC-GFMI}. This suggests that typical \gls{RMS} models used in commercial software more accurately represent the behavior of a \gls{VC-GFMI} than that of a \gls{CC-GFMI}.

\begin{figure}[b]
    \centering
    \includegraphics[width=\linewidth]{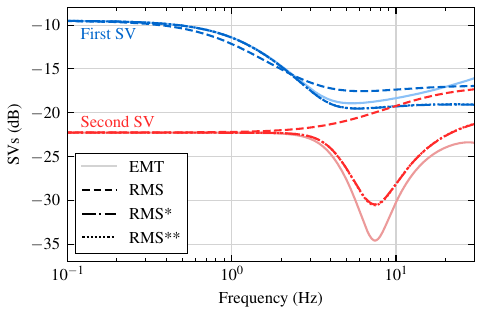}
    \caption{Low-frequency scans of $\dqMatrix{Z}(s)$ at the \gls{PCC} for the \gls{EMT} and different \gls{RMS} models of the \gls{CC-GFMI}.}
    \label{fig:GFM_improved}
\end{figure}

To further investigate the observed mismatch for the \gls{CC-GFMI} case, we first include the inner voltage-control loops (RMS*) and, as a second step, the inner current-control loops (RMS**). The results are shown in Fig.~\ref{fig:GFM_improved} for the frequency range from $\qtyrange{0.1}{30}{\Hz}$. Including the inner voltage-control loops significantly improves the accuracy of the RMS scan at low frequencies, whereas adding the inner current-control loops additionally has a negligible effect, and is, in any case, infeasible for \gls{RMS} simulations \cite{ExaminingFeasibilityModeling2023Mohiuddin}. This is expected since the \gls{EMT} model of the \gls{CC-GFMI} controls the terminal voltage $\dqVector{u}$, while the \gls{RMS} model controls the internal voltage $\dqVector{v}$. By including the inner voltage-control loops in the \gls{RMS} model, it also controls the terminal voltage $\dqVector{u}$ and therefore aligns with the \gls{EMT} model.

\subsection{Time-Domain Validation}
To confirm that \gls{RMS} simulations hide potential high-frequency instabilities, we perform a time-domain simulation with the \gls{GFLI} connected to the three-bus system. We consider a poorly-tuned \gls{PLL}, i.e. increasing its proportional gain, and apply a step change of $\qty{0.125}{\pu}$ in the active power reference at $t=\qty{0.5}{\second}$, followed by a step change of $\qty{-0.01}{\pu}$ in the terminal voltage reference at $t=\qty{1.5}{\second}$.

\begin{figure}
    \centering
    \includegraphics[width=\linewidth]{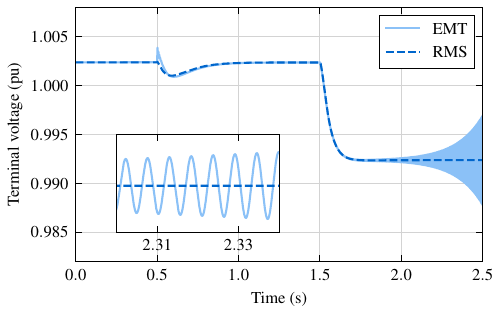}
    \caption{Time-domain simulation of \gls{GFLI} in three-bus system with poorly-tuned \gls{PLL}.}
    \label{fig:GFL_TD}
\end{figure}

As seen in Fig.~\ref{fig:GFL_TD}, the \gls{EMT} simulation shows instability with an oscillation frequency of approximately $\qty{190}{\Hz}$ after the step change in the voltage reference. However, the system remains stable in the \gls{RMS} simulation, confirming that \gls{RMS} models cannot capture high-frequency instabilities. The oscillation frequency observed for the \gls{EMT} model closely matches the frequency of the unstable mode of the linearized system and one of the peaks observed in the first \gls{SV}.

\subsection{Interpretation of EMT Frequency Scans}
As the \gls{RMS} models cannot capture high-frequency dynamics, we now limit our analysis to the \gls{EMT} frequency scans.

\begin{figure}
    \centering
    \includegraphics[width=\linewidth]{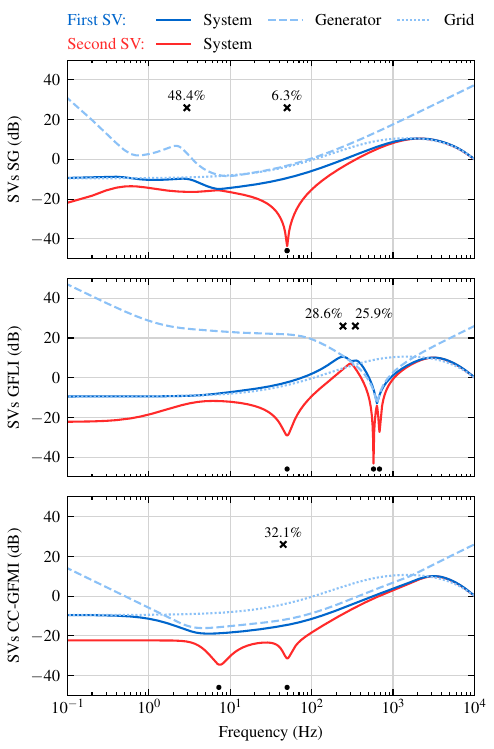}
    \caption{Frequency scans of the first \gls{SV} of $\dqMatrix{Z}(s)$, $\dqMatrixSup{Z}{gen}(s)$, and $\dqMatrixSup{Z}{grid}(s)$ for the \gls{SG}, \gls{GFLI}, and \gls{CC-GFMI}, as well as of the second \gls{SV} of $\dqMatrix{Z}(s)$. Oscillatory modes of the system with a damping ratio below 50~\% are indicated by black crosses with their damping ratios in $\%$. Selected zeros are indicated by black dots.}
    \label{fig:EMT_interpretation}
\end{figure}

First, we investigate the relation between $\dqMatrix{Z}(s)$, $\dqMatrixSup{Z}{gen}(s)$, and $\dqMatrixSup{Z}{grid}(s)$ using the first \gls{SV}. In Fig.~\ref{fig:EMT_interpretation}, the first \glspl{SV} of these three impedances are shown for the \gls{SG}, \gls{GFLI}, and \gls{CC-GFMI}. As derived in Section~\ref{sec:Impedance_Scan}, the lower of $\dqMatrixSup{Z}{gen}(s)$ and $\dqMatrixSup{Z}{grid}(s)$ dominates $\dqMatrix{Z}(s)$. Hence, at very low and high frequencies, $\dqMatrixSup{Z}{grid}(s)$ dominates, which is the same for the three generator types. Conversely, the generator influence is strongest in the frequency range from $\qtyrange{3}{1000}{\Hz}$ for the \gls{SG}, from $\qtyrange{500}{3000}{\Hz}$ for the \gls{GFLI}, and from $\qtyrange{0.5}{3000}{\Hz}$ for the \gls{CC-GFMI}. At high frequencies, the influence of $\dqMatrixSup{Z}{gen}(s)$ becomes negligible, since it increases due to the inductance of the transformer and the stator in the case of the \gls{SG}. The same principle applies to the second \gls{SV}. However, the generator influence is strongest in different frequency ranges.

Second, we connect the frequency-scan results to the small-signal analyses of the linearized system models. The curves of interest are the first and second \glspl{SV} of $\dqMatrix{Z}(s)$ in Fig.~\ref{fig:EMT_interpretation}. Due to space limitations, we summarize the most significant findings qualitatively. The modes of the linearized models reveal that each peak observed in the frequency scan of the first \gls{SV} of $\dqMatrix{Z}(s)$ can be associated with an oscillatory mode, marked by a black cross and labeled with its damping ratio in Fig.~\ref{fig:EMT_interpretation}. The height of the peak depends on the residue, i.e., the product of the mode's observability and controllability, as well as on the inverse of the magnitude of the real part of the oscillatory mode. Dips in the second \gls{SV} are associated with zeros of the system, marked by black dots in Fig.~\ref{fig:EMT_interpretation}. When there is a pole and a zero at the same or a similar frequency, the oscillatory mode can be hidden, as is the case for the \gls{SG} and the \gls{CC-GFMI} at around 50~\si{Hz}. Note that this is different from a pole-zero cancellation, because the real parts of the pole and the zero differ. Since the locations of the zeros depend on the selected inputs and outputs, modes hidden in the frequency scan of the system impedance at one location might be visible in a frequency scan from another location in the system.

\section{Conclusion}
\label{sec:conclusions}

This work shows that \gls{RMS} models in frequency scans capture low-frequency dynamics up to \qty{10}{\Hz}, but they can mask potential high-frequency interactions between the generator and the grid. The typical \gls{RMS} model of a \gls{GFMI} captures the low-frequency behavior of a \gls{VC-GFMI}. However, inaccuracies arise for a \gls{CC-GFMI} even at low frequencies. The mismatch can be reduced by including the inner voltage-control loops into the \gls{RMS} model of the \gls{CC-GFMI}. Additionally, we show that the system impedance is dominated by the lower of the grid and the generator impedances. The peaks observed in the frequency scan correspond to oscillatory modes of the system. However, certain oscillatory modes might be masked by zeros at the same frequency. Future work will investigate the operating point sensitivity of EMT frequency scans.

\bibliographystyle{IEEEtran}
\bibliography{references.bib}

\end{document}